\documentstyle[prl,aps,multicol,psfig]{revtex}
\begin{document}
\draft
\title{Spin accumulation in ferromagnetic
single-electron transistors \\ in the cotunneling regime}
\author {Jan Martinek$^{1,2}$, J\'ozef Barna\'s$^{1,3}$,
 Sadamichi Maekawa$^2$,
 Herbert Schoeller$^{4,5}$, and Gerd Sch\"{o}n$^{4,6}$}
\address{$^1$Institute of Molecular Physics, Polish Academy of Sciences,
 ul. Smoluchowskiego 17, 60-179 Pozna\'n, Poland\\
 $^2$Institute for Materials Research, Tohoku University, Sendai
980-8577, Japan\\
 $^3$Department of Physics, Adam Mickiewicz
University, ul. Umultowska 85, 61-614 Pozna\'n, Poland\\
 $^4$Forschungszentrum Karlsruhe, Institut f\"ur Nanotechnologie,
76021 Karlsruhe, Germany \\
 $^5$Institut
f\"{u}r Theoretische Physik A, RWTH Aachen, 52056 Aachen, Germany
\\
 $^6$Institut f\"{u}r Theoretische Festk\"{o}perphysik,
 Universit\"{a}t Karlsruhe, 76128 Karlsruhe, Germany \\
}

\date{Received \hspace{5mm} }
\date{\today}
\maketitle

\begin{abstract}
We propose a new method of direct detection of spin accumulation,
which overcomes problems of previous measurement schemes. A spin
dependent current in a single-electron transistor with
ferromagnetic electrodes leads to spin accumulation on the
metallic island. The resulting spin-splitting of the
electrochemical potentials of the island, because of an additional
shift by the charging energy, can be detected from the spacing
between two resonances in the current-voltage characteristics. The
results were obtained in the framework of a real-time diagrammatic
approach which allows to study higher order (co-)tunneling
processes in the strong nonequlibrium situation.
\end{abstract}

\begin{multicols}{2}

The discovery of materials and devices with spin dependent
electronic transport properties opens new perspectives for
research and applications. As examples we mention exchange-coupled
metallic magnetic multi-layers, heterostructures with
ferromagnetic semiconductors, hybrid structures based on magnetic
metals  and nonmagnetic semiconductors, as well as perovskite
oxides. The field has developed into a new branch of mesoscopic
electronics -- called magneto- or spin electronics \cite{prinz}.
The balance of spin injection and spin diffusion in spin-polarized
heterostructures in general leads to a non-equilibrium spin
accumulation \cite{aronov,valet}, which in turn produces a
difference in the local electrochemical potentials for spin-up and
-down electrons. The predicted spin polarization was probed in
experiments, either by measuring the resistance of metallic
devices \cite{johnson,wees} or in semiconducting systems
\cite{kikkawa} by studying the circular polarized
photoluminescence.

In spite of the reported successes in the indirect detection of
spin accumulation, there remain controversies about the results
\cite{wees}. In these methods one usually measures the resistance
and for a typical experimental configuration some other effects
related to interface scattering, anisotropic magnetoresistance,
etc, can occur.
 Even the question whether in principle it is
possible to observe the spin splitting of the electrochemical
potential by spectroscopic methods (analogous to the tunneling
spectroscopy for the superconducting gap) remains open. The reason
is the fact that the spin accumulation is a nonequilibrium effect,
and the splitting of electrochemical potential is always smaller
than the applied bias (in contrast to the superconducting energy
gap which does not depend on the applied voltage).

In this article, we propose a method of direct detecting the
nonequilibrium spin splitting of the electrochemical potentials in
ferromagnetic single-electron transistors (FM SETs),
 which is free of the
difficulties met in the methods used previously.
 We show that
the splitting can be read from the spacing between two resonances
in the current-voltage characteristics. The resonances are
separated due to the spin dependence of the electrochemical
potentials, but in addition they are shifted by the charging
energy due to Coulomb blockade effects.

Spin-dependent transport in ferromagnetic double-barrier tunnel
junctions has been studied before, both experimentally \cite{ono}
and theoretically in the sequential
\cite{barnas1,brataas,martinek}, cotunneling
\cite{maekawa1,maekawa2}
 and strong tunneling \cite{wang} regimes. The previous work covering
the cotunneling regime is based on the approach developed by
Averin and Nazarov \cite{nazarov}
 for nonmagnetic SETs,
  which, however, is
valid only away from the resonance. On the other hand, exactly the
resonance is crucial for the effects considered in this paper.
Therefore, we extend the previous descriptions by using the
diagrammatic real-time technique, developed for normal-metal SETs
\cite{schoeller,konig}, and extend it to FM SETs with
spin-dependent tunneling. We take into account single-barrier
cotunneling processes \cite{nazarov1}, vertex correction and
propagator renormalization.

The system under study consists of a small nonmagnetic  metallic
grain (island) connected to two ferromagnetic leads (electrodes)
 via tunnel barriers \cite{brataas,martinek,maekawa2}.
Its Hamiltonian takes the form
\begin{equation}
\label{hamiltonian}
        H=\sum_{\rm r=L,R} H_{\rm r} + H_{\rm I}+H_{\rm
ch}+H_{\rm T} \equiv H_0+H_{\rm T}\; .
\end{equation}
Here $H_{\rm r}=\sum_{k \nu\sigma}\epsilon^{\rm r}_{k \nu\sigma}
a^\dagger_{{\rm r}k \nu\sigma} a_{{\rm r}k \nu\sigma}$ (for
$r=L,R$) and $H_{\rm I}=\sum_{q\nu\sigma}\epsilon_{q\nu\sigma}
c^\dagger_{q\nu\sigma} c_{q \nu\sigma}$ describe noninteracting
electrons
in the two leads and island, respectively. The eigenstates of the
leads and island are described by the wave-vectors $k$ and $q$,
transverse channel index $\nu$, and spin $\sigma$. The Coulomb
interaction on the island is accounted for by $H_{\rm ch}=E_{\rm
C} (\hat{n}-n_x)^2$, where $E_{\rm C}=e^2/2C$ is the scale for the
charging energy, and $C=C_{\rm L}+C_{\rm R}+C_{\rm g}$ is the
total capacitance of the island, which is the sum of the left and
right junction capacitances and the gate capacitance.
 The `external charge' $en_x \equiv C_{\rm L} V_{\rm L}+C_{\rm R} V_{\rm
R}+C_{\rm g} V_{\rm g}$ accounts for the effect of the applied
voltages, $V_{\rm L}$ and $V_{\rm R}$ in the left and right
electrodes and the gate voltage $V_{\rm g}$.
 The last part of the Hamiltonian, $H_{\rm T}$, describes tunneling
processes and may be written as
\begin{equation}
\label{tunneling}
        H_{\rm T}=\sum_{\rm r=L,R}\sum_{kq}\sum_{\nu\sigma}
        T^{{\rm r}\sigma}_{kq\nu }
        a^\dagger_{{\rm r}k \nu\sigma}c_{q \nu\sigma}{\rm e}^{-i\hat{\varphi}}+
{\rm
        h.c.} \, .
\end{equation}
The phase operator $\hat{\varphi}$ is the conjugate to the charge
 $e\hat{n}$ on the island, and the operator $e^{\pm i\hat{\varphi}}$
 describes changes of the island charge by $\pm e$.

When writing Eq.~(\ref{tunneling}) we assumed that the electron
spin and transverse channel index are conserved during tunneling.
We further will assume for simplicity that the transfer matrix
elements $T^{{\rm r}\sigma}_{kq \nu}$  are only
dependent on the junction $r$ and the spin orientation $\sigma$,
$T^{{\rm r}\sigma}_{kq \nu}=T^{\rm r}_\sigma$. They can be related
to the spin dependent tunneling resistance of the barriers via the
relation
 $1/R_{{\rm r}\sigma}=(2 \pi e^2/\hbar )
   N \left|T^{{\rm r}}_{\sigma}\right|^2
     D_{{\rm r}\sigma} D_{\rm I}$,
where $D_{{\rm r}\sigma}$ is the spin-dependent density of
electron states at the Fermi level in the electrode $\rm r$,
$D_{\rm I}$ the density of states per spin in the island, and $N$
the number of transverse channels.

We first ignore intrinsic spin-flip processes in the island, since
usually the spin flip time $\tau_{\rm sf}$ is much longer than the
time scale for the variation of the charge on the island and the
energy relaxation time. Thus, we treat the two spin components of
the electrons on the island as two independent electron
reservoirs. They are characterized by Fermi distributions $f_{\rm
I \sigma}(E) = 1/\{ \exp [\beta (E - \mu_{\rm I\sigma})] + 1\}$,
where $\mu_{\rm I \sigma}$ are the corresponding spin dependent
electrochemical potentials which are different in a general case,
$\mu_{\rm I \uparrow} \neq \mu_{\rm I \downarrow}$
\cite{macdonald}. Without loss of generality we can choose the
reference energy such
 that they satisfy $\mu_{\rm I \uparrow} = - \mu_{\rm I \downarrow}$.
Spin current conservation, as described later, fixes these values.

The grand-canonical density matrix of the system depends on the
electrochemical potentials of the two spin components. Following
the standard procedure \cite{schoeller,konig} we expand it in
$H_{\rm T}(t)$ and perform the trace over the reservoir degrees of
freedom using Wick's theorem.
The electric current is given in the lowest order perturbation
expansion by $I^{(1)} = \sum_{\sigma} I^{(1)}_{\rm L \sigma} =
-\sum_{\sigma} I^{(1)}_{\rm R \sigma}$, with
\begin{eqnarray}
\label{eq:current1}
        I^{(1)}_{\rm r \sigma} &=& {4 \pi^2 e\over h}\sum_n
        \left[p_n^{(0)}+p_{n+1}^{(0)}\right] \nonumber \\
         && \times
        {\alpha^-(\Delta_n) \, \alpha_{\rm r \sigma}^+(\Delta_n)-
        \alpha^+(\Delta_n) \, \alpha_{\rm r \sigma}^-(\Delta_n) \over
        \alpha(\Delta_n)} \; .
\end{eqnarray}
Here $\Delta_n=E_{\rm ch}(n+1)-E_{\rm ch}(n)$
, and $\alpha^\pm_{\rm r \sigma}(\epsilon )$ are the forward and
backward propagators on the Keldysh contour in Fourier space,
\begin{eqnarray}
\label{eq:alpha3}
         \alpha^\pm_{\rm r \sigma}(\epsilon ) = \pm \alpha^0_{\rm r \sigma}
          {\epsilon -\Delta \mu_{\rm r \sigma}
        \over \exp[\pm \beta(\epsilon -\Delta \mu_{\rm r \sigma})]-1}\; .
\end{eqnarray}
The dimensionless conductance of the junction $r$ for spin
$\sigma$ is $\alpha^0_{\rm r \sigma} = h/(4 \pi^2 e^2 R_{\rm r
\sigma})$, $\epsilon$ is the energy of the tunneling electron, and
$\Delta\mu_{\rm r \sigma}=\mu_{\rm r}-\mu_{\rm I \sigma}$.
Apart from this, $\alpha^\pm(\epsilon) = \sum_{\rm r \sigma}
\alpha_{\rm r, \sigma}^\pm (\epsilon)$, $\alpha (\epsilon)=
\alpha^+(\epsilon) +\alpha^-(\epsilon)$, and the probabilities
$p_n^{(0)}$ obey the equation $p_n^{(0)} \alpha^+(\Delta_n) -
p_{n+1}^{(0)} \alpha^-(\Delta_n) = 0$ with $\sum_n p_n^{(0)}=1$.

The dominant second order (cotunneling) contribution to the
electric current can be divided into three parts, $I^{(2)} =
\sum_{i=1}^3 I^{(2)}_i$, with
 $I^{(2)}_i = \sum_{\sigma} I^{(2)}_{i
\rm L \sigma} =- \sum_{\sigma} I^{(2)}_{i \rm R \sigma}$. These
terms are given by the following equations:
\begin{eqnarray}
\label{eq:curren2}
        I^{(2)}_{1 \rm r \sigma} &=& {4 \pi^2 e\over h}
        \sum_n p_n^{(0)} \int d\omega \,
        \big[ \alpha^-(\omega) \,
        \alpha_{\rm r \sigma}^+(\omega) \, \nonumber \\ &&
        - \, \alpha^+(\omega) \,
        \alpha_{\rm r \sigma}^-(\omega) \big]
         \,
        {\rm Re} \, R_n(\omega)^2 \, , \\
        I^{(2)}_{2 \rm r \sigma} &=& - {1\over 2} {4 \pi^2 e\over h}
        \sum_n \left( p_n^{(0)}+p_{n+1}^{(0)}\right) \nonumber \\ &&
        \times
        {\alpha^-(\Delta_n) \, \alpha_{\rm r \sigma}^+(\Delta_n)-
        \alpha^+(\Delta_n) \, \alpha_{\rm r \sigma}^-(\Delta_n)
        \over \alpha(\Delta_n)} \nonumber \\ &&
        \times \int d\omega \, \alpha(\omega) {\rm Re}
        \left[ R_n(\omega)^2 + R_{n+1}(\omega)^2 \right] \, , \\
        I^{(2)}_{3 \rm r \sigma} &=& - {1\over 2} {4 \pi^2 e\over h}
        \sum_n  \left( p_n^{(0)}+p_{n+1}^{(0)}\right) \nonumber \\ &&
        \times
        {\partial \over \partial \Delta_n} \,
        \left[
        {\alpha^-(\Delta_n) \, \alpha_{\rm r \sigma}^+(\Delta_n)-
        \alpha^+(\Delta_n) \, \alpha_{\rm r \sigma}^-(\Delta_n) \over
        \alpha(\Delta_n)} \right] \nonumber  \\ &&
        \times \int d\omega \, \alpha(\omega) {\rm Re} \,
        \left[ R_n(\omega) - R_{n+1}(\omega) \right] \, ,
\end{eqnarray}
where
      $  R_n(\omega)=1 /( \omega-\Delta_n+i0^+) -
        1 /( \omega-\Delta_{n-1}+i0^+) \,$.
The terms $I_2^{(2)}$ and $I_3^{(2)}$ describe renormalization of
the tunneling conductance and energy gap, respectively, and become
important at resonance.

We have numerically evaluated the current for parallel (P) and
antiparallel (AP) alignment of the electrode magnetizations. We
assumed symmetric junctions such that for parallel alignment
 $R_{\rm R
\uparrow} = R_{\rm L \uparrow} = a R_{\rm Q}$
and  $R_{\rm R \downarrow} = R_{\rm L \downarrow} =
(1-P)/(1+P)R_{\rm R \uparrow}$, where
 $R_{\rm Q} = h/e^2$
and  we assume $a = 5$. The parameter $P$ denotes the spin
polarization of the electrodes -
$P=0.23$ and
$0.40$ for $\rm Ni$ and $\rm Fe$, respectively. For the
antiparallel alignment one then finds $R_{\rm R \uparrow} = R_{\rm
L \downarrow} = a R_{\rm Q}$, $R_{\rm R \downarrow} = R_{\rm L
\uparrow} = (1-P)/(1+P)R_{\rm R \uparrow}$.

The spin accumulation on the island (or equivalently spin
splitting of the electrochemical potential) is determined from the
condition of spin current conservation
\begin{eqnarray}
\label{eq:spincurrent}
        \sum_{\rm r}
        ( I_{\rm r \sigma}^{(1)} + I_{\rm r \sigma}^{(2)}
        )- e\mu_{\sigma}D_{\rm I}/\tau_{\rm sf} = 0        \, .
\end{eqnarray}
Here the spin flip processes in the island are taken into account,
and characterized by the relaxation time $\tau_{\rm sf}$.

Equation~(\ref{eq:curren2}) includes the components,
which describe single-barrier cotunneling processes which
transport spin but do not transport charge. They cancel after
summation over the spin orientation and do not enter the current
formula. On the other hand, they enter Eq.~(\ref{eq:spincurrent})
and, by modifying the magnetic state of the island, have an
indirect influence on the transport current.

\begin{figure}
\centerline{\psfig{figure=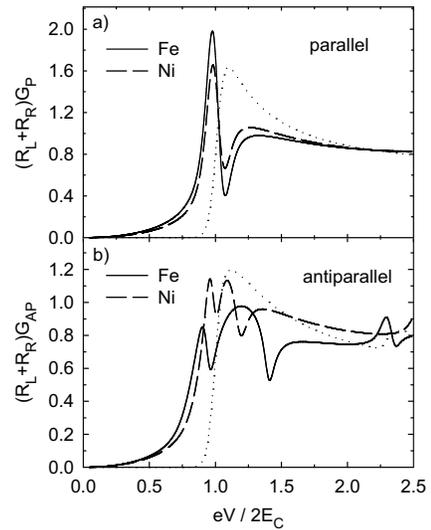,width=5.5cm}}
 \caption{Differential conductance vs.\ bias voltage $V$ in the (a)
parallel and (b) antiparallel configurations, calculated for
$T/E_{\rm C}=0.02$ and symmetric junctions with $R_{\rm R
\uparrow} = 5 R_{\rm Q}$. The spin polarization is $P=$ 0.23 and
0.40 for $\rm Ni$ and $\rm Fe$ electrodes, respectively. Dotted
line is for the sequential tunneling limit for $\rm Fe$
electrodes.}
 \label{fig1}
\end{figure}

\begin{figure}
\centerline{\psfig{figure=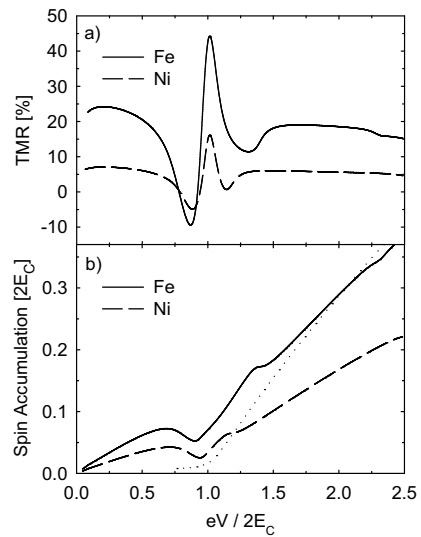,width=5.4cm}}
 \caption{(a) Tunnel magnetoresistance $\rm TMR$ and (b) spin
 accumulation for
$\rm Ni$ and $\rm Fe$ electrodes as functions of bias voltage $V$.
The other parameters are the same as in Fig.~\ref{fig1}.
 Dotted line is for the sequential tunneling limit for $\rm Fe$
electrodes.}
 \label{fig2}
\end{figure}

In Fig.~\ref{fig1} we show the differential conductance for $\rm
Ni$ and $\rm Fe$ electrodes in both parallel and antiparallel
magnetic configurations, calculated for $T/E_{\rm C}=0.02$ in the
absence of spin-flip processes. The corresponding value of tunnel
magnetoresistance, ${\rm TMR} \equiv(R_{\rm AP}-R_{\rm P})/R_{\rm
P}$
  where $ R_{\rm P} $ and $ R_{\rm AP} $ are total resistances for parallel
  and antiparallel configuration respectively,
  is shown in Fig.~\ref{fig2}(a). In the AP configuration the
splitting of the conductance peak, resulting from the spin
splitting of the electrochemical potential of the island, is well
resolved (see Fig.~\ref{fig1}(b)). Different splitting for Fe and
Ni is due to different magnetic polarizations $P$ of these metals.

 The magnitude of the associated
spin accumulation is shown in Fig.~\ref{fig2}(b). It is
interesting to note that no spin accumulation, and consequently no
splitting of the resonance, occurs in the P configuration (see
Fig.~\ref{fig1}(a)).

\begin{figure}
\centerline{\psfig{figure=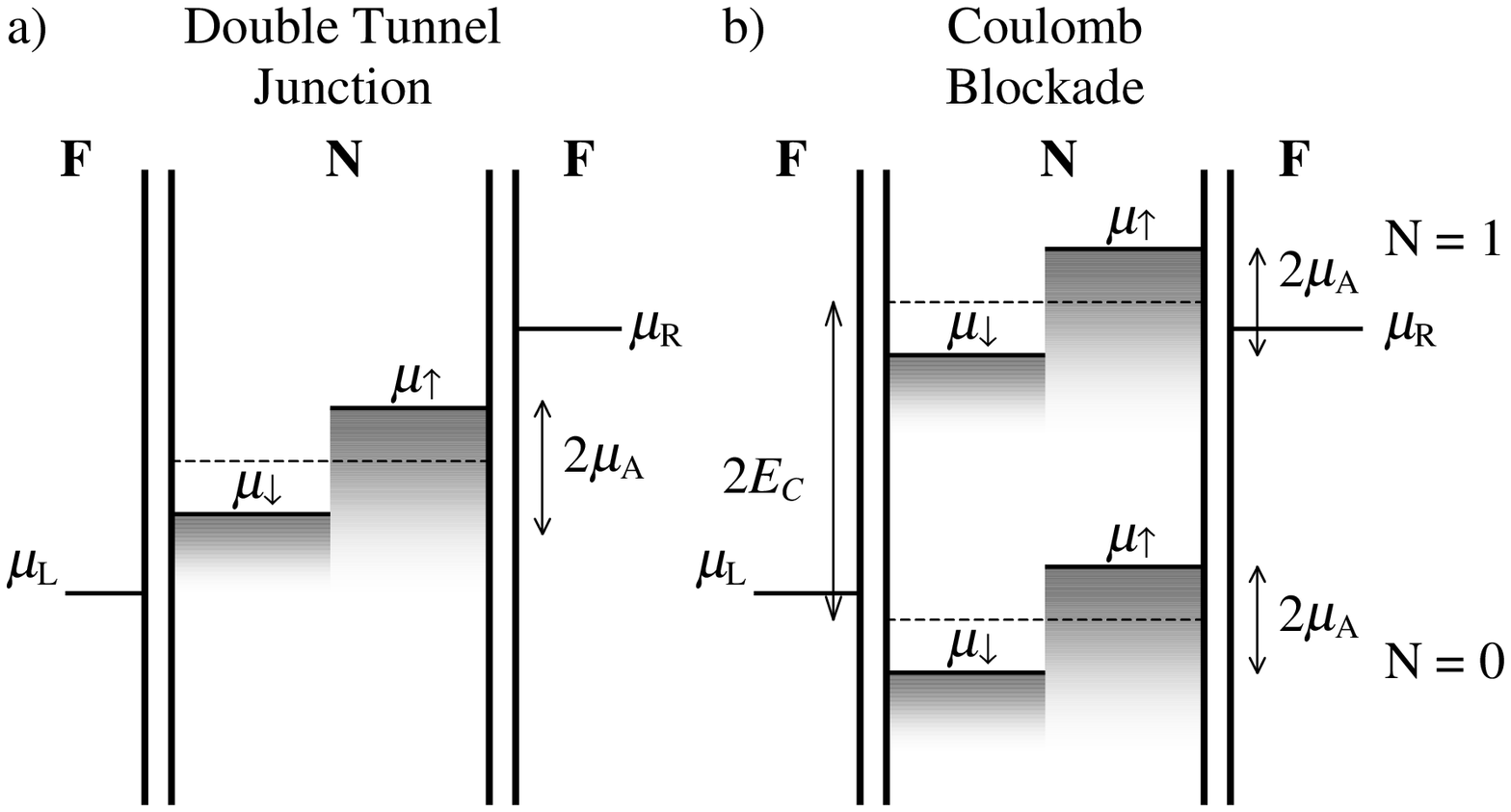,width=7cm}}
 \caption{Energy
diagrams for a symmetric magnetic double tunnel junction with a
normal metallic island for the antiparallel configuration (a)
without and (b) with the Coulomb blockade. Here $\mu_{\rm L}$,
$\mu_{\rm R}$ are the electrochemical potentials for the left and
right electrodes and $\mu_{\uparrow}$, $\mu_{\downarrow}$ are the
electrochemical potentials for spin-up and spin-down electrons on
the island. $N=1 \, ,0$ denotes the island charge.}
 \label{fig3}
\end{figure}

The cotunneling component of the current shows resonances when the
bias voltage $V$ approaches the - effective - electrochemical
potential of either spin component. The pure spin splitting of the
electrochemical potentials is always smaller then the bias
voltage. However, Coulomb blockade effects effectively shift the
electrochemical potentials further by the charging energy (as
illustrated in the diagram of Fig.~\ref{fig3}). Due to this
additional shift the resonances in the current should be
resolvable. A direct observation of the resonances in the current
may still be difficult because even at resonance cotunneling
yields only a small fraction of the total current. The resonances
are easier to detect in the differential conductance, which is
shown in Fig.~\ref{fig1}. For the interpretation of these results
it should be noted that a resonance in the current corresponds to
a maximum negative slope of the conductance.

Such an effect may also arise in the sequential tunneling limit
\cite{martinek1} (high resistance junctions), but in that case it
is not so evident and pronounced as in the cotunneling limit (see
Fig.~\ref{fig1}). The role of the cotunneling current is twofold:
(i) the cotunneling current has a sharp maximum at resonance
(described above), which can be easily detected on the conductance
characteristics (see comparison of cotunneling and sequential
limit in Fig.~\ref{fig1}), and (ii) in the Coulomb blockade regime
sequential tunneling is suppressed, so only the cotunneling
processes contribute to spin accumulation (as can be concluded
from Fig.~\ref{fig2}(b) for $eV/2E_{\rm C} < 1$).

\begin{figure}
\centerline{\psfig{figure=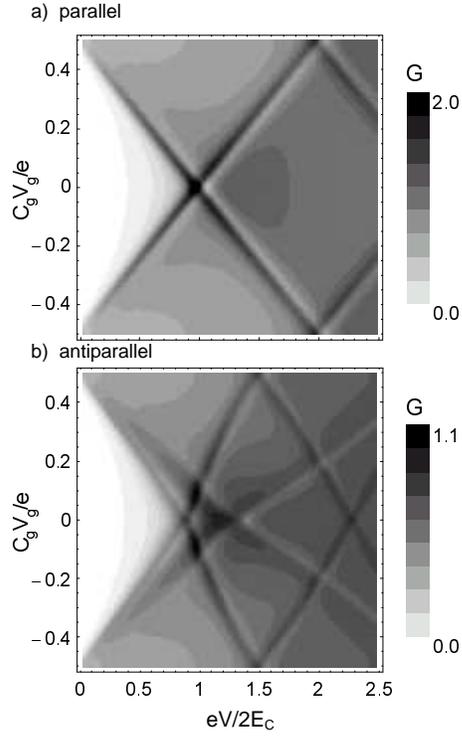,width=6cm}}
 \caption{The differential conductance vs.\ gate $V_{\rm g}$ and
  transport $V$ voltages in a gray-scale representation
  in the (a) parallel and (b) antiparallel configurations
  calculated for $\rm Fe$ electrodes and
   the other parameters as in Fig.~\ref{fig1}.}
 \label{fig4}
\end{figure}

In Fig.~\ref{fig4} we show the differential conductance versus
gate $V_{\rm g}$ and transport $V$ voltages in a gray-scale plot
for both P and AP magnetic configurations. The splitting of the
resonance (shown in Fig.~\ref{fig1}(b) for $V_{\rm g}=0$) is
clearly visible in Fig.~\ref{fig4}(b). From this figure one can
read out the dependence of the electrochemical potential splitting
on the transport voltage $V$. Figure ~\ref{fig4}(b) also shows
that the splitting of the conductance peaks can be observed when
the gate voltage $V_{\rm g}$ is varied and the bias voltage $V$ is
fixed.

\begin{figure}
\centerline{\psfig{figure=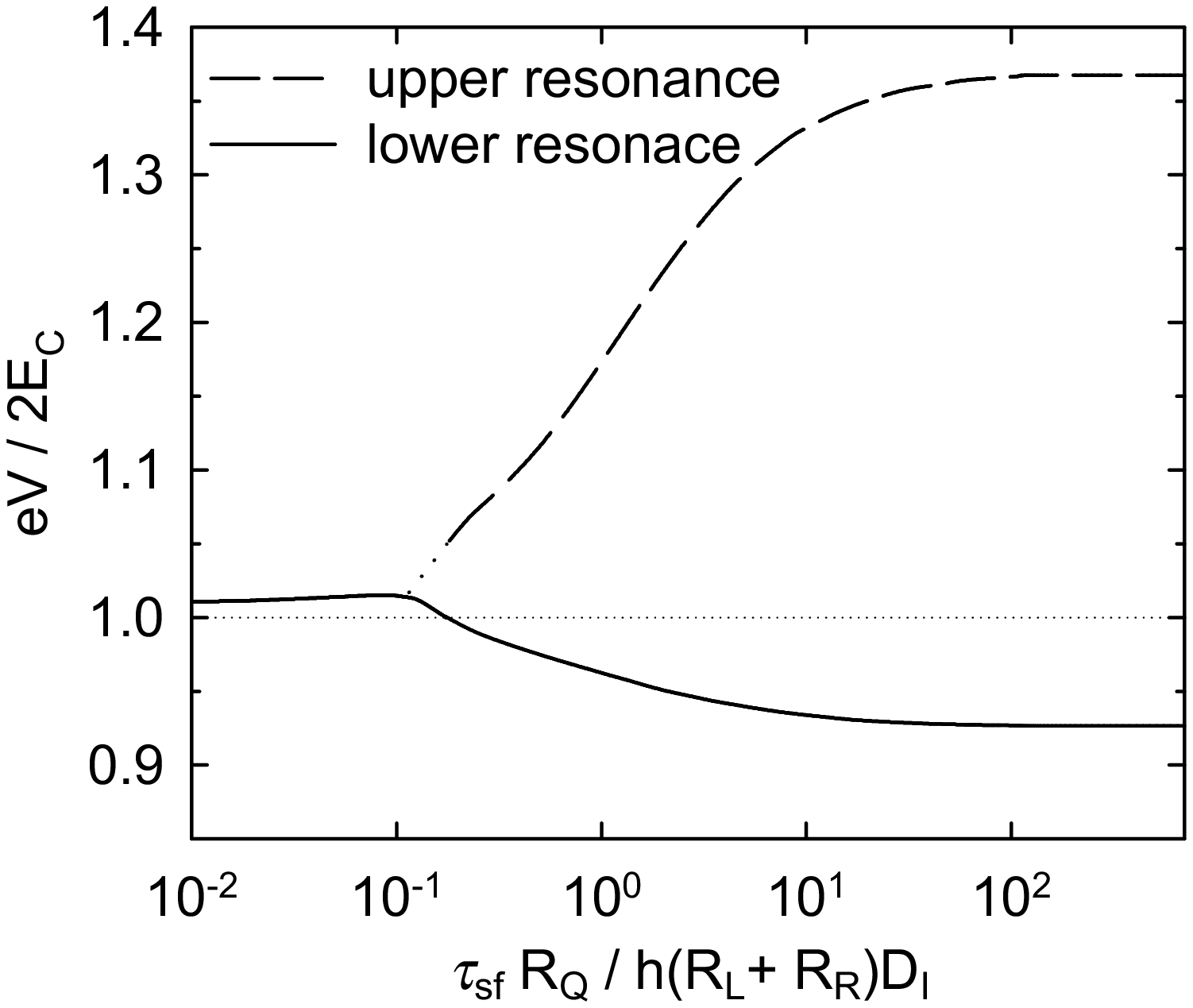,width=5.0cm}}
 \caption{Dependence of the upper and lower resonance voltages
  on the intrinsic spin
 relaxation time $\tau_{\rm sf}$, calculated for $\rm Fe$
  electrodes and the other parameters as in Fig.~\ref{fig1}.
  The dotted line denotes a regime of insufficiently resolved resonances.}
 \label{fig5}
\end{figure}

Generally, spin-flip relaxation processes suppress spin
accumulation.
 Figure~\ref{fig5} shows the values of the voltage
corresponding to the upper and lower resonances
 from Fig.~\ref{fig1}(b)
(maximum negative slopes of the conductance curves) as a function
of the intrinsic spin relaxation time $\tau_{\rm sf}$ on the
island. Generally, the spin-relaxation time $\tau_{\rm sf}$
depends on sample properties. For instance, in $\rm{Al}$ one finds
$\tau_{\rm sf} \sim 10^{-8} s$. For small islands, $D_{\rm I}
\simeq 10^3/eV$, and for the other parameters as in
Fig.~\ref{fig1}, one finds $\tau_{\rm sf} R_{\rm Q}/h(R_{\rm L} +
R_{\rm R} )D_{\rm I} \sim 10^3$. This means that the spin
relaxation process is relatively slow and the value of the spin
accumulation is close to its maximum value for $\tau_{\rm sf}
\rightarrow  \infty $. It is known that spin accumulation can be
suppressed by a small perpendicular magnetic field (`Hanle
effect') \cite{johnson}. In principle, this effect can be measured
in the system under consideration
   and may be used to determine the spin relaxation time as well.

 There are several experimental works \cite{ono} on ferromagnetic SETs
 both in the sequential and cotunneling regime
with all electrodes (including island) ferromagnetic. The
corresponding data, as well as the data obtained on ferromagnetic
granular systems suggest that spin accumulation may occur in such
systems, but there was no clear experimental evidence on this. But
the range of the parameters indicates a possibility of a direct
detection by the method proposed in this paper.

In conclusion, we have analyzed electron tunneling in
ferromagnetic SETs in the cotunneling regime. We showed that the
spin accumulation on the island can be observed directly in the
current-voltage characteristics as a splitting of the resonance.
The distance between two resonances arises due to the
spin-splitting of the electrochemical potential due to spin
accumulation.
 This effect should be observable
experimentally, and could give clear evidence of the spin
accumulation, overcoming the objections to the previous
measurements \cite{johnson,wees}.

The paper is supported by the Polish State Committee for
Scientific Research under the Project No. 5 P03B 091 20 and the
EU-COST program action P5.
 One of us (J.M.) would like to
acknowledge the hospitality of the University and
Forschungszentrum Karlsruhe, where part of this work was
performed.

\end{multicols}
\end{document}